\documentclass[]{spie}  

\usepackage{amsmath,amsfonts,amssymb}
\usepackage{graphicx}
\usepackage[colorlinks=true, allcolors=blue]{hyperref}
\usepackage{siunitx}
\usepackage{placeins}
\usepackage{multicol}
\usepackage{multirow}
\usepackage{array}
\usepackage{booktabs}
\usepackage{setspace}
\usepackage{tocloft}
\usepackage{subcaption}
\usepackage{wrapfig}
\usepackage{makecell}

\title{Continued developments in X-ray speed reading: fast, low noise readout for next-generation wide-field imagers}

\author[a]{Sven Herrmann}
\author[a]{Peter Orel}
\author[a]{Tanmoy Chattopadhyay}
\author[a]{Glenn Morris}
\author[b]{Gregory Prigozhin}
\author[a,e]{Haley R. Stueber}
\author[a,d,e]{Steven W. Allen}
\author[b]{Marshall W. Bautz}
\author[b]{Kevan Donlon}
\author[b]{Beverly LaMarr}
\author[c]{Chris Leitz}
\author[b]{Eric Miller}
\author[a,e]{Abigail Pan}
\author[a]{Artem Poliszczuk}
\author[a]{Daniel R. Wilkins}

\affil[a]{Kavli Institute of Astrophysics and Cosmology, Stanford University, 452 Lomita Mall, Stanford, CA 94305, USA}
\affil[b]{Kavli Institute for Astrophysics and Space Research, Massachusetts Institute of Technology, Cambridge, MA, USA}
\affil[c]{MIT Lincoln Laboratory, Lexington, MA, USA}
\affil[d]{SLAC National Accelerator Laboratory, 2575 Sand Hill Road, Menlo Park, CA 94025, USA}
\affil[e]{Department of Physics, Stanford University, 382 Via Pueblo Mall, Stanford CA 94305, USA}

\authorinfo{Further author information: (Send correspondence to Sven Herrmann)\\Sven Herrmann: E-mail: svench@stanford.edu}

\begin{document} 
\maketitle

\begin{abstract}
Future strategic X-ray astronomy missions will require unprecedentedly sensitive wide-field imagers providing high frame rates, low readout noise and excellent soft energy response. To meet these needs, our team is employing a multi-pronged approach to advance several key areas of technology. Our first focus is on advanced readout electronics, specifically integrated electronics, where we are collaborating on the VERITAS readout chip for the Athena Wide Field Imager, and have developed the Multi-Channel Readout Chip (MCRC), which enables fast readout and high frame rates for MIT-LL JFET (junction field effect transistor) CCDs. Second, we are contributing to novel detector development, specifically the SiSeRO (Single electron Sensitive Read Out) devices fabricated at MIT Lincoln Laboratory, and their advanced readout, to achieve sub-electron noise performance. Hardware components set the stage for performance, but their efficient utilization relies on software and algorithms for signal and event processing. Our group is developing digital waveform filtering and AI methods to augment detector performance, including enhanced particle background screening and improved event characterization. All of these efforts make use of an efficient, new X-ray beamline facility at Stanford, where components and concepts can be tested and characterized. 
\end{abstract}


\section{INTRODUCTION}
\label{sec:intro}  
\FloatBarrier

Future strategic X-ray astronomy missions \cite{AXIS2023, Athena17} will require fast, sensitive, wide-field imagers. High frame rates will be essential to minimize the impact of pile-up on measurements of point sources, and of the particle background on studies of faint, diffuse gas. At the same time, these imagers will require excellent soft energy response and correspondingly low noise to enable their full discovery potential. The depleted p-channel field effect transistor (DEPFET) detectors being developed for the Athena Wide Field Imager (WFI)\cite{NOM_WFI_2016} will deliver excellent performance. However, the large pixel size of this technology makes it best suited to missions maximizing collecting area and readout speed over spatial resolution \cite{LAZI_DEPFET_2022}. DEPFET detectors also carry substantial readout microelectronics requirements \cite{NOM_WFI_2016}. State-of-the-art X-ray charge-coupled devices (CCDs), in contrast, are available with small pixels and are capable of delivering all of the key performance metrics for missions seeking the best possible spatial resolution, like AXIS\cite{ericspie2023} and Lynx\cite{2019JATIS...5b1001G}, except for the required frame rates. Our group at Stanford is addressing these technology gaps via a multi-pronged approach, in collaboration with colleagues at the Massachusetts Institute of Technology (MIT), MIT Lincoln Laboratory (MIT-LL) and the Max Planck Institute for Extraterrestrial Physics (MPE).  

\begin{figure} [ht]
   \begin{center}
   \begin{tabular}{c}
   \includegraphics[height=6.5cm]{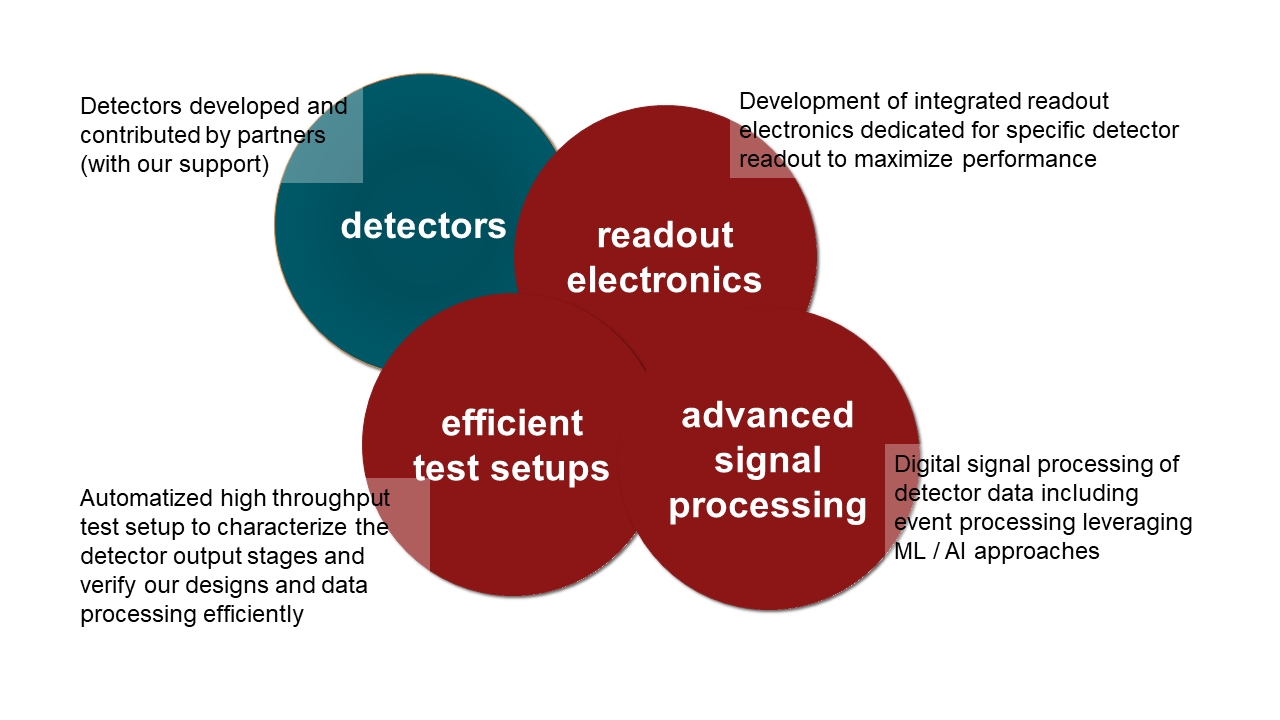}
   \end{tabular}
   \end{center}
   \caption[example] 
   {\label{fig:block} 
To fill the technology gap for future X-ray missions we employ a multi-prong development strategy that advances detectors, electronics and signal processing including X-ray event processing.}
   \end{figure}

\section{(Integrated) Readout Electronics}
One of the key areas of development to close the technology gap for future X-ray observatories is readout electronics for the large and fast X-ray imagers these missions require. In order to achieve the required fame rates while keeping (or even improving) the excellent noise performance of state of the art X-ray CCDs, the pixel readout rate (expressed in MPix/s) needs to be increased by 10 to 100 times. The key roadmap for this is to increase the number of readout nodes for a given imager, while also increasing the speed per readout node. The best path to implement this is to develop dedicated integrated electronics due to their low parasitic capacitance, small footprint and low power consumption, a critical feature when many channels need to be accommodated. 

\FloatBarrier
\subsection{VERITAS readout ASIC for Athena WFI}
\label{sec:VERITAS}

\begin{figure} [ht]
   \begin{center}
   \begin{tabular}{c}
   \includegraphics[height=4cm]{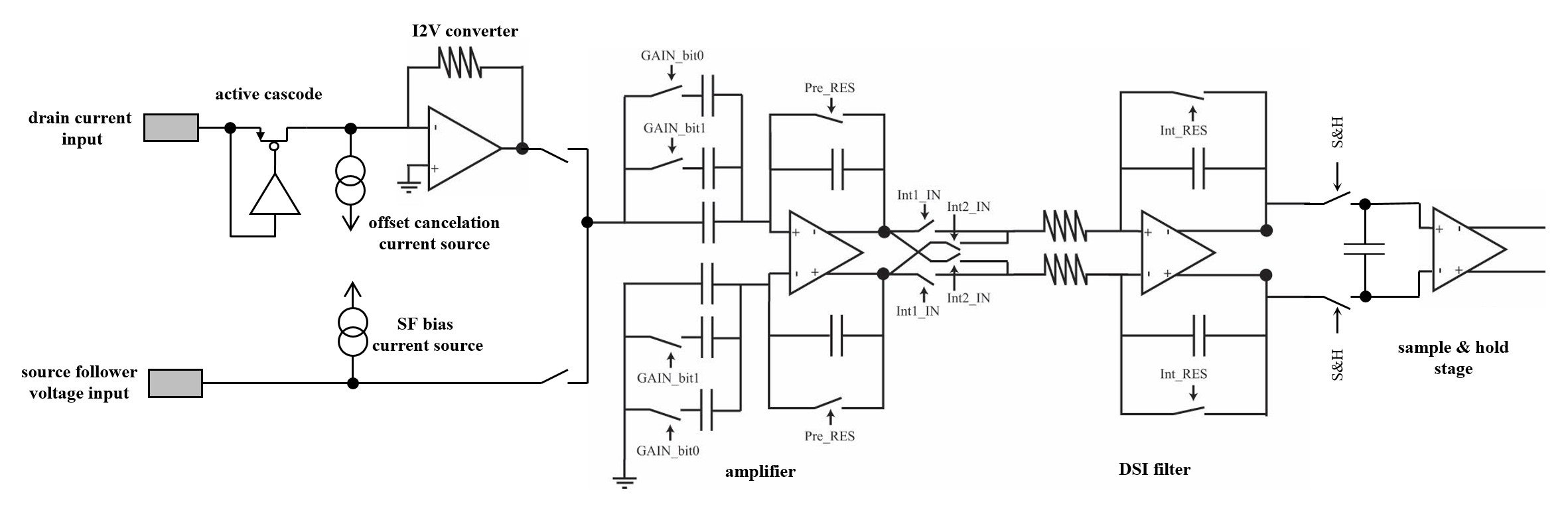}
   \end{tabular}
   \end{center}
   \caption[example] 
   {\label{fig:VERITAS_channel} 
   Overview schematic of a VERITAS 2.x readout channel including the main components of a single analog processing cell. The full chip integrates 64 of these channels operating in parallel.}
   \end{figure}

As part of the US contribution to the European Space Agency (ESA) Athena mission, our group has pioneered drain current readout architectures (in contrast to conventional source follower voltage readout) for the DEPFET based wide field imager (WFI) and we work collaboratively with the Max Planck Institute for Extraterrestrial Physics on the readout application specific integrated circuit (ASIC) for the DEPFET detector. Drain readout does not suffer from the settling time limitation as conventional source follower based readouts, which allows for an increase in speed per output - in the case of the Athena WFI DEPFET detector this speedup is roughly a factor of two (5 \unit{\micro s} VS 2.5 \unit{\micro s}). In addition the, the VERITAS readout ASIC integrates a large number (64) of channels operating in parallel enabling the full column parallel readout of the WFI DEPFET detector, achieving a WFI full frame rate of 500 frames per second. Figure \ref{fig:VERITAS_channel} shows the overview schematic of one single VERITAS 2 readout channel: it includes the two inputs, one for source follower voltage readout and one for drain current readout, a differential preamplifier with user selectable gain setting, followed by a dual slope integrator (DSI) filter and a sample \& hold (S\&H) stage. After the S\&H stage all 64 channels are multiplexed to the output through a single unity gain fully differential output buffer.
Further details can be found in Porro et al. [\citenum{Porro_VERITAS2_2014}] and Schweingruber et al. 2024 [\citenum{SchweingruberSPIE2024}]. The VERITAS 2 targets a per channel readout rate of 2.5 \unit{\micro s} or 400 kHz - at these speeds the analog DSI filter is a particularly good choice as the detector waveforms are well behaved and can readily be integrated. The integration time of around 1 \unit{\micro s} is a good match for the available practical implementations of resistance and capacitance values into the chip.

\subsection{Multi-channel CCD Readout Chip for a future X-ray mission}
\label{sec:MCRC}

\begin{figure}
    \centering
    \begin{subfigure}{.5\textwidth}
    \includegraphics[width=\linewidth]{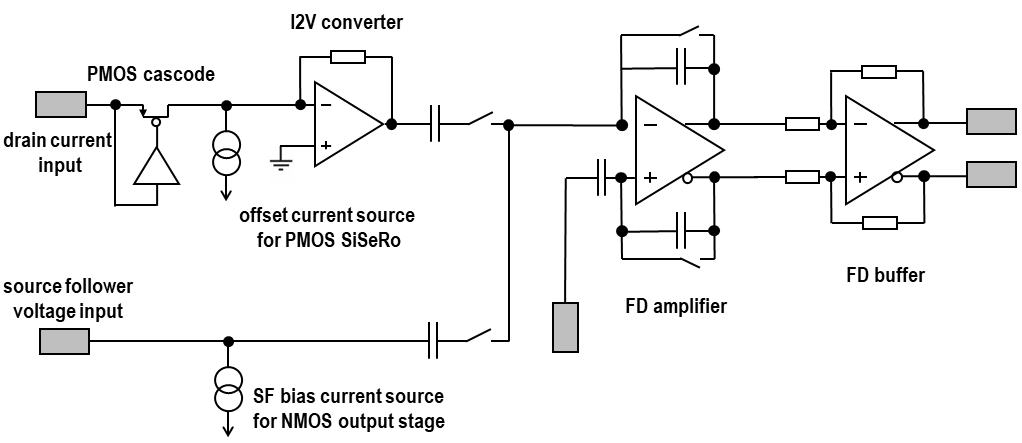}
    \caption{}
    \end{subfigure} \hfill
    \begin{subfigure}{.4\textwidth}
    \includegraphics[width=\linewidth]{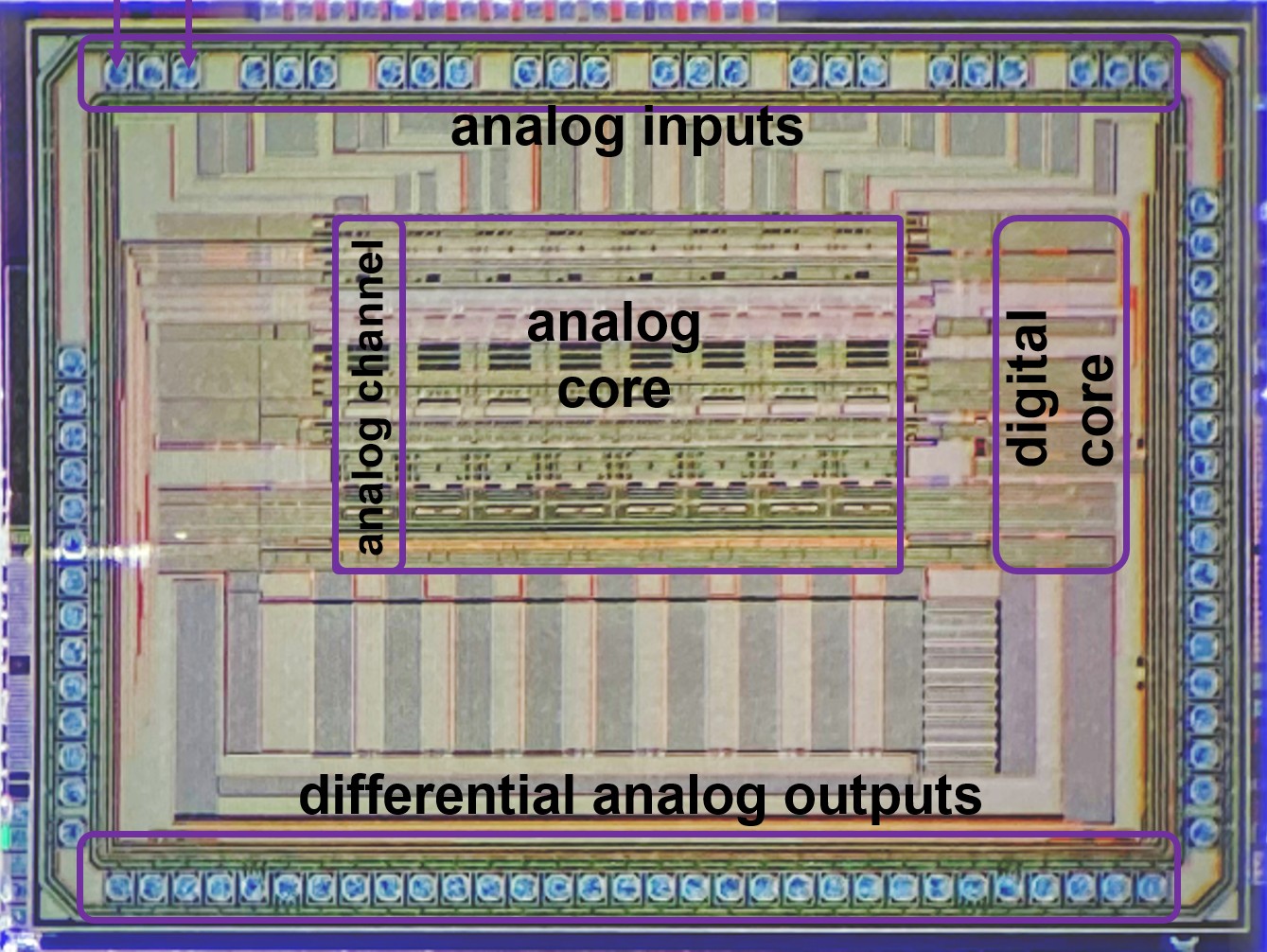}
    \caption{}
    \end{subfigure} \hfill
    \caption{(a) Overview schematic of the MCRC analog channel. Each channel consists of an input stage, which is select-able between voltage input and current input and provides the bias for the CCD output stage, a preamplifier for signal amplification, and a fully differential buffer driver that drives the signal to an external ADC. (b) a micro-photograph of the MCRC V1.0 ASIC. The 8 channel chip is approx. 4.2 mm by 3.0 mm large with the analog inputs shown on the top side and the outputs on the bottom side. Pads on on the right side are used for the digital interface, while the pads on the right side are for debugging and bias control.}
    \label{fig:MCRCV1_pic}
\end{figure}

The MCRC is an integrated analog readout ASIC, manufactured in a 3.3V 0.35 \unit{\micro m} process technology, designed and optimized for the readout of MIT-LL X-ray CCDs with JFET or SiSeRo outputs. While a later iteration of the ASIC could include more channels, the MCRC-V1 features 8 analog readout channels operating in parallel.  Each of the analog channels, as shown in figure \ref{fig:MCRCV1_pic}, consists of an input stage, which is selectable between voltage input and current input and provides the bias for the CCD output stage, a preamplifier for signal amplification and single-ended to differential conversion, and a fully differential unity-gain output buffer driver. The amplified video waveform is sampled by an external analog-to-digital converter (ADC) after which digital pulse processing is used to extract the image data. In addition, the chip includes a digital serial peripheral interface (SPI) to program the ASIC settings and control the internal switch logic and bias digital-to-analog converters (DACs) to set the operating points of the internal amplifiers, and logic to generate internal clock signals from an external master clock. The ASIC provides the performance (speed and noise) of our best discrete readout amplifier implementation in an integrated package with a fraction of the footprint and power consumption. While the overall structure bears resemblance with the VERITAS readout channel the distinct difference is the achievable speed, which for MCRC is around 5 MPix/s and as such more than 10x higher then that of VERITAS. This not only requires much faster amplifiers, but also a different approach for the signal filtering as is outlined in section \ref{sec:AdvDSP}. Details of the performance evaluation can be found in Orel et al. 2022 [\citenum{porelMCRCspie2022}] and 2024 [\citenum{porelMCRCspie2024}]. Figure \ref{fig:MCRCV1_pic} shows a micro-photograph of the manufactured ASIC which is approx 4.2 mm by 3.0 mm in size. Together, these features and capabilities make the readout ASIC the ideal readout electronics solution for future large and fast X-ray imagers like the proposed Advanced X-ray Imaging Satellite (AXIS) high-speed X-ray camera. The high functional integration of the readout ASIC reduces not only the footprint but also the number of components, significantly simplifying the CCD board design.

\FloatBarrier
\section{Detector Test Beamline}

\begin{figure} [ht]
   \begin{center}
   \begin{tabular}{c}
   \includegraphics[height=7.8cm]{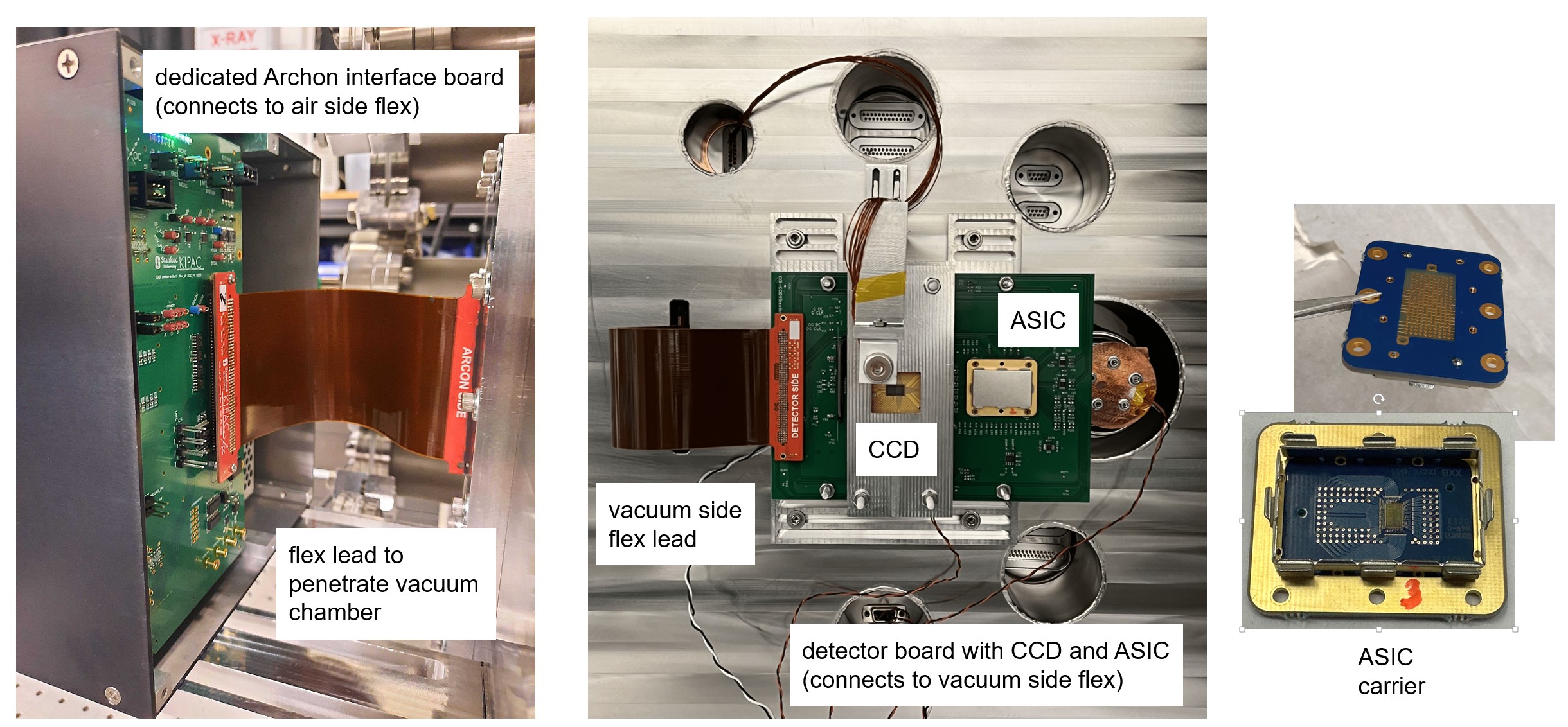}
   \end{tabular}
   \end{center}
   \caption[example] 
   {\label{fig:beamline_boards} 
   Detector electronics of the test beamline: STA Archon controller to control the detector and sample the signal waveforms (left); detector PCB with readout ASIC and flex lead (middle); MCRC ASIC carrier to house the readout chip (right) }
   \end{figure}

To test and characterize our microelectronics and advanced signal processing efforts, we have developed and built an efficient test beamline consisting of a large vacuum chamber to house the X-ray imager and an X-ray tube with fluorescence targets connected to the chamber via a 2.5 meter long beamline. This arrangement results in a reasonable homogeneous illumination suitable for testing. Detailed information about the beamline design and its performance is given in Stueber et al. [\citenum{stueberSPIE2024}]. 
While the detector and the front end readout electronics are located inside the vacuum, we use a commercial CCD readout controller on the air side of the chamber to provide the clock signals and digitize the analog video waveform. A flex lead printed circuit board is potted into a vacuum flange and connects the vacuum printed circuit board (PCB) with the air side readout controller. Figure \ref{fig:beamline_boards} shows two photographs of the setup with the air and vacuum side.  

The vacuum side PCB includes not only the detector but also the preamplifier and offers the option of using either a discrete electronics implementation or the MCRC-V1 ASIC for the readout. To maximize versatility, we decided not to wire bond the readout ASIC to the CCD directly (which would have been best for performance), instead we placed the readout ASIC onto its own small carrier PCB that connects to the detector board via a Samtex Z-stack connector. These connectors provide low parasitic capacitance and are well suited for this application. The detector itself is also easily changeable, using a PGA ZIF socket (Pin Grid Array Zero Insertion Force socket). Such a connection scheme enables us to change the detector or readout chip easily, in order to evaluate device to device variations or to replace the device in case of failure. In addition, having both the discrete readout implementation and the readout ASIC on the board, facilitates the comparison between the two. 

The amplified detector video-waveform is transmitted via the flex lead to the CCD readout controller for digitization and processing. For this we use a Semiconductor Technology Associates, Inc (STA\footnote[5]{http://www.sta-inc.net/archon/}) FPGA-based instrument dubbed the STA Archon, that provides control signals like clocks and biases to the detector and simultaneously digitizes the CCD video signal. In addition, the Archon controller features an internal digital signal processing pipeline that extracts the relevant information from the acquired detector waveform to reconstruct a 2-dimensional image. For our application at the test beamline we have enhanced the commercial controller with a custom interface board that includes buffers, voltage regulators and debugging features. A more detailed description of the Archon controller modules is given in [\citenum{chattopadhyay20}].

\subsection{MCRC V1.0 + CCID-93 X-ray results }
\label{sec:title}

\begin{figure}
   \begin{center}
   \begin{tabular}{c}
   \includegraphics[height=6cm]{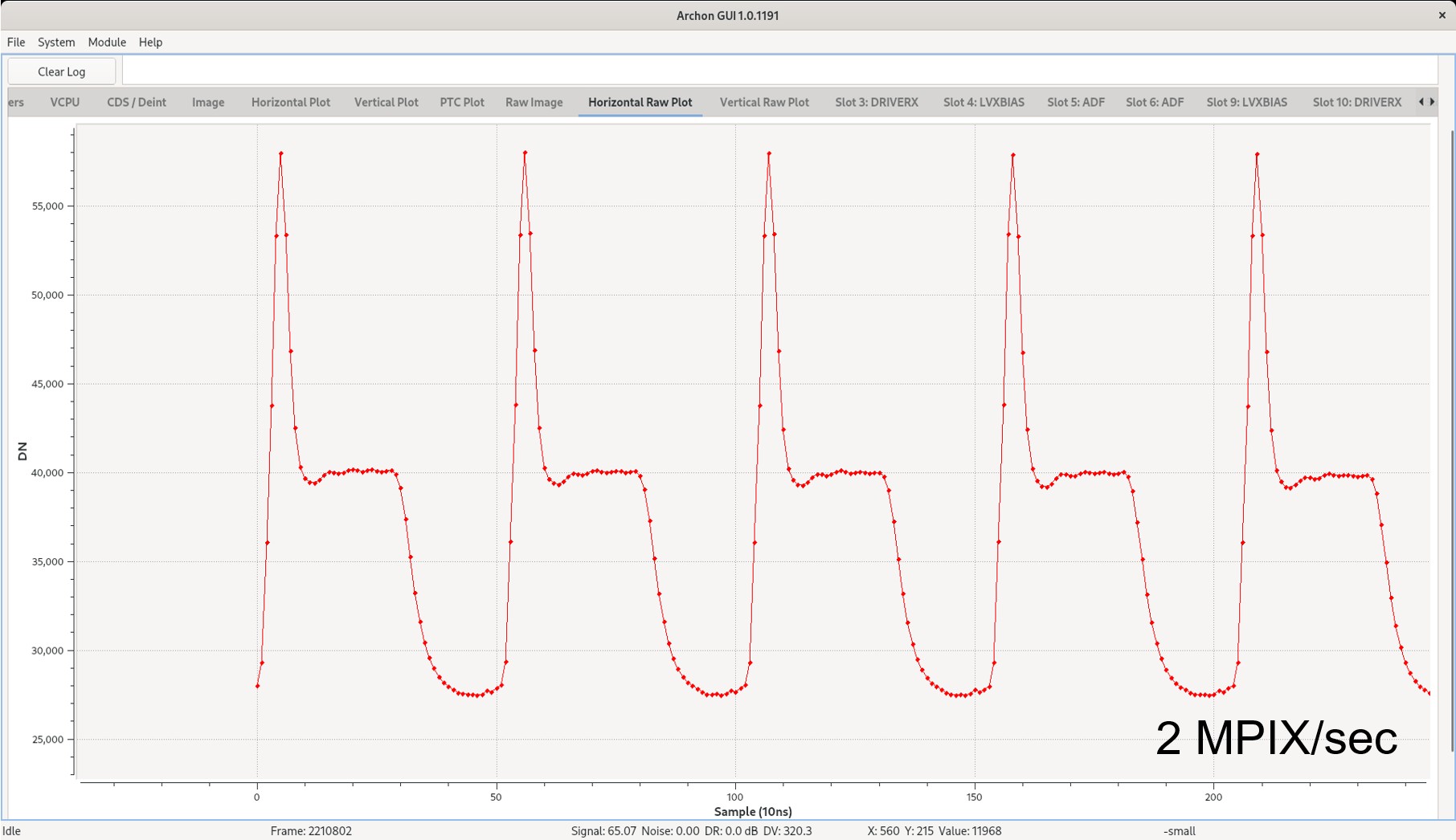}
   \end{tabular}
   \end{center}
   \caption[example] 
   {\label{fig:MCRC_2MPIX_videowaveform} 
Screenshot from the STA Archon CCD controller depicting the 2 MPix/s video-waveform captured by the CCID-93 JFET CCD read out with the MCRC-V1 readout ASIC. The large spikes upwards are the output stage reset, followed by the baseline level and the settling of the signal level. }
\end{figure}

\begin{figure} [ht]
   \begin{center}
   \begin{tabular}{c}
   \includegraphics[height=7cm]{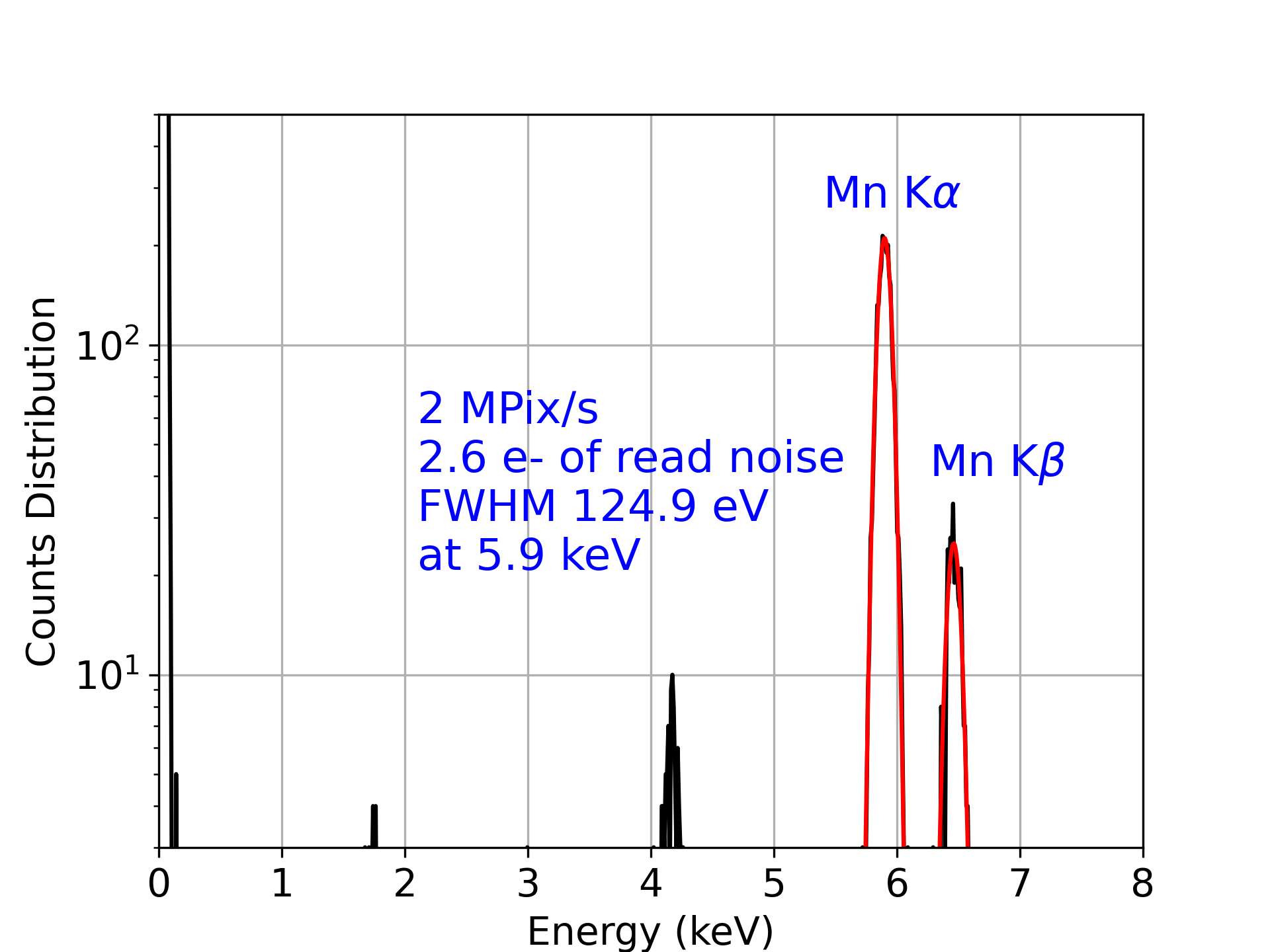}
   \end{tabular}
   \end{center}
   \caption[example] 
   {\label{fig:MCRC_Xray_spectrum} 
X-ray measurement of MCRC-V1 paired with a MIT-LL CCID93 pJFET based X-ray CCD with an Iron-55 ($^{55}$Fe) source at 2 MPix/s per output. The captured spectrum shows the excellent result with a measured read noise of 2.6 \unit{e^{-}_{RMS}} and a resulting ($^{55}$Fe) spectrum with a FWHM of 125 eV.}
\end{figure}

After the evaluation of the MCRC V1.0 readout ASIC without an detector by means of electronic parameter tests \cite{porelMCRCspie2022} and the detailed characterization of the CCID-93 X-ray CCD with a discrete amplifier \cite{ericspie2023}, we finally combined the components in our new beamline for tests and characterization of the CCID-93 \& MCRC-V1 combination. In these measurements the MCRC readout ASIC provides the biasing to the JFET output stage of the detector, so that no further external components are needed - the CCD output can be directly connected to the readout ASIC, significantly simplifying the system design. The CCD was cooled to -90\unit{\degree C} and we utilized a pixel rate of 2 MPix/s, which is our default rate to obtain the best characterization. The resulting video waveform as captured by the Archon controller can be seen in Figure \ref{fig:MCRC_2MPIX_videowaveform} - the shape of this waveform is dominated by the CCD output stage and as such identical to the response of the discrete amplifier implementation. More information about the comparison of MCRC-V1 readout ASIC with an established discrete amplifier implementation can be found in Orel et al. 2024 [\citenum{porelMCRCspie2024}]. The measured read noise was around 2.6 \unit{e^{-}_{RMS}}, virtually identical to the discrete alternative and the resulting Iron-55 ($^{55}$Fe) spectrum as shown in Figure \ref{fig:MCRC_Xray_spectrum} exhibits a FWHM of 124.9 eV - again virtually identical to the discrete alternative.


\FloatBarrier
\section{Advanced Signal Processing}
\label{sec:AdvDSP}

Simultaneously with our efforts on readout electronics we also identified advanced signal processing as a necessity to close the performance gaps for future missions.
While the VERITAS readout ASIC for the Athena WFI DEPFET detector uses an integrated DSI (dual slope integrator) filter, this efficient implementation comes to its limits when the readout rates increase substantially. For future applications we are targeting rates up to 5 MPix/s per output, a 10x increase from the VERITAS 2.3 0.5 MPix/s rate.
Therefore, the MCRC concept does not rely on an analog signal filter to separate the measured X-ray signal from the noise, but employs video waveform sampling with digital signal processing. This approach has gotten more attention in recent years and such commercial options for digital video waveform  processing like the STA Archon controller we use are available on the market. 
The STA Archon also allows for raw waveform export and processing. Our group demonstrated noise improvements by optimizing the signal processing specifically for output stages with high 1/f noise contributions. Details and results have been published in Chattopadhyay et al. 2023 [\citenum{chattopadhyay23_sisero_digfilt}]  
Recently we focused our efforts onto Repetitive Non-Destructive Readout (RNDR) techniques : methods which measure the same charge signal multiple times in order to converge onto a smaller measurement error - potentially reaching deep sub-electron noise performance that can count individual electrons. Not every detector output stage is suitable for that but in recent years a number of different devices have started to exploit this method with success \cite{tiffenberg17,RNDR_DEPFET_2007} including SiSeRO devices \cite{sofo23_nsisero}.

\subsection{SiSeRO RNDR - subelectron read noise}
\label{sec:SISERO_RNDR}
The SiSeRO readout stage development led by MIT Lincoln Laboratory (MIT-LL) is a novel charge detector concept for X-ray CCDs, with a working principle similar in some respects to the DEPFET sensors \cite{kemmer87_depfet,strueder00_depfet_imager} developed for the Athena WFI \cite{treberspurg20_wfi}, which is drawing on earlier work on floating-gate amplifiers that demonstrated extremely high responsivity and sub-electron \cite{matsunaga91} noise. The SiSeRO amplifier comprises a MOSFET straddling the transfer-channel of a CCD’s output register. When a charge packet is transferred beneath the MOSFET channel, it modulates the transistor drain current, which can be sensed directly with a current amplifier. SiSeRO devices can be built with either a P-MOS transistor that senses electrons, or an N-MOS transistor that senses charge packets constituted from holes \cite{sofo23_nsisero}. The P-MOS SiSeRO is fully compatible with the established MIT-LL X-ray CCDs, prompting our efforts to focus on this device flavor. 
In Chattopadhyay et al. 2022 [\citenum{chattopadhyay22_sisero}], we used the first manufactured SiSeRO devices to demonstrate their working principle, and presented results using a drain current readout module. 

The SiSeROs also offer the feature for repetitive non-destructive readout (RNDR): since the charge packet in the internal gate is unaffected by the readout process, it can be moved around like any charge packet in a CCD and therefore measured multiple times. This process can reduce the noise below the 1/f barrier, deep into the sub-electron regime. This capability not only improves the detector performance in the important soft-X-ray regime, but also could lead to simplified gain calibration procedures \cite{rodrigues21}, with the gain of the electronic chain able to be determined at the level of individual electrons. 

\begin{figure} [ht]
   \begin{center}
   \begin{tabular}{c}
   \includegraphics[height=7.5cm]{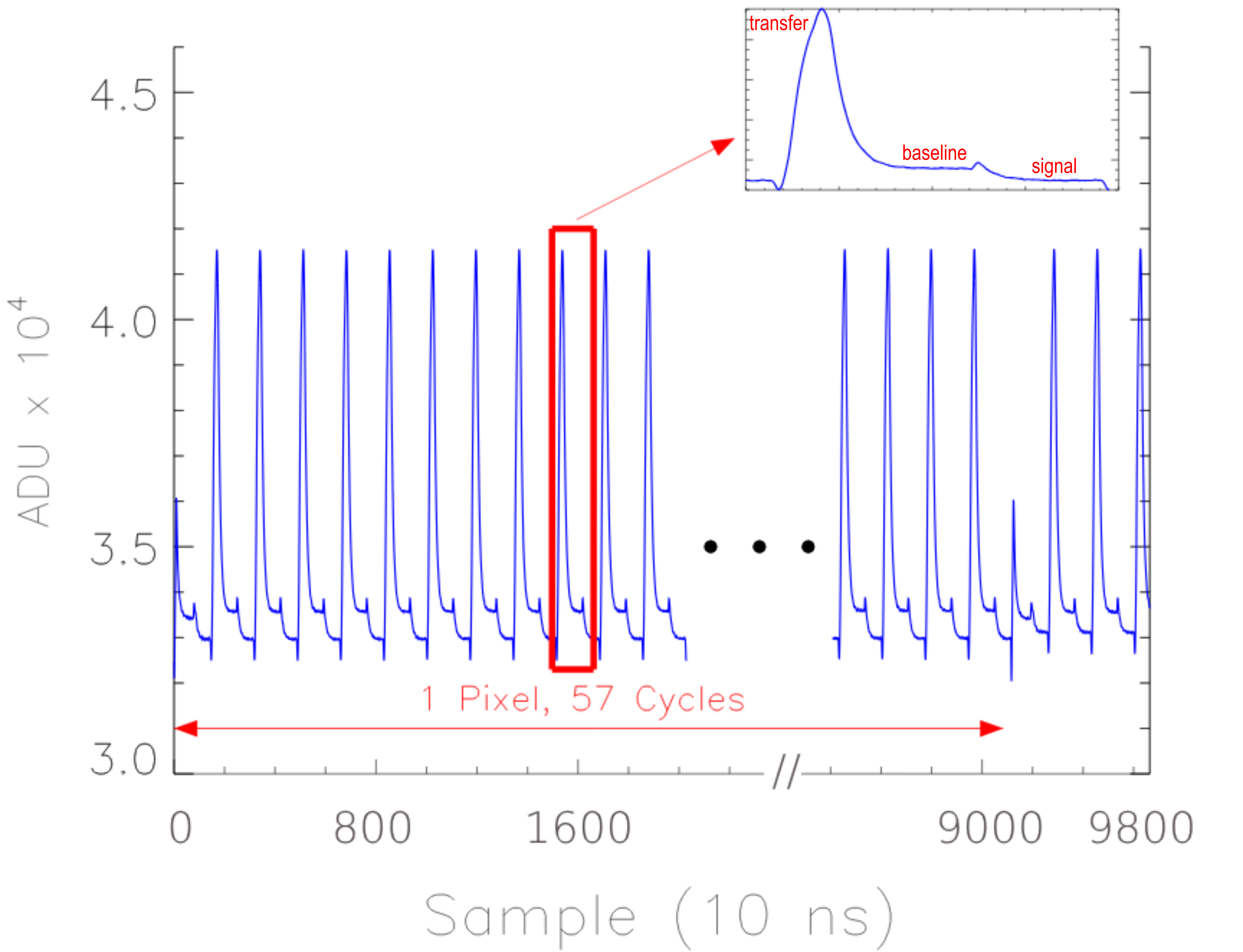}
   \end{tabular}
   \end{center}
   \caption[example] 
   {\label{fig:RNDR_27_waveform} 
RNDR measurement with SiSeRO devices: single pixel waveform captured for 57 RNDR cycles. The large spikes upwards are from the clocking of the adjacent SiSeRO gates to pull the charge back out from the internal gate, while the actual signal is the change from empty internal gate to filled internal gate.}
\end{figure}

\begin{figure} [ht]
   \begin{center}
   \begin{tabular}{c}
   \includegraphics[height=7cm]{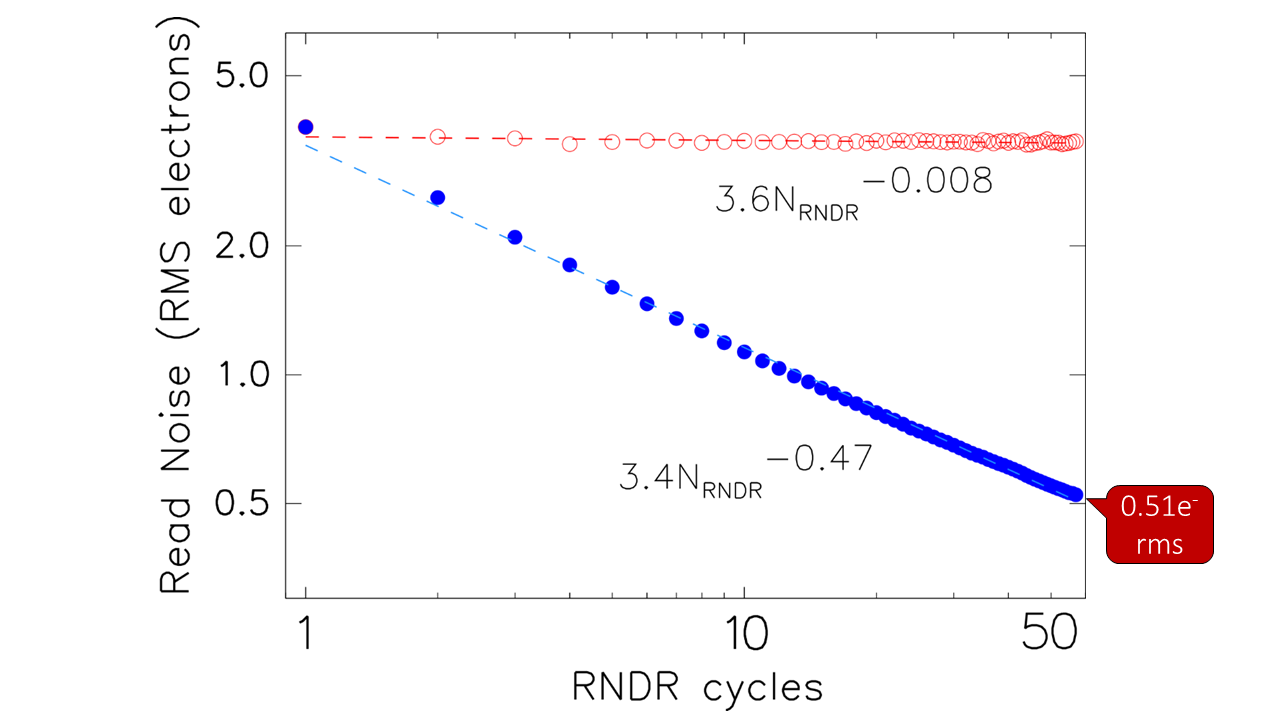}
   \end{tabular}
   \end{center}
   \caption[example] 
   {\label{fig:RNDR_27_noise} 
RNDR measurement with SiSeRO devices: noise performance vs number of RNDR cycles. The red circles represent the single cycle read noise measurements, while the blue circles denote the read noise of the sequentially (cycle-by-cycle) compounded measurements. After 13 cycles, the ($\sim$20 \unit{\micro s}) 1.0 \unit{e^{-}_{RMS}} readout noise threshold is reached, while 57 RNDR cycles ($\sim$90 \unit{\micro s}) lead us to achieve a readout noise of 0.5 \unit{e^{-}_{RMS}}.}
\end{figure}

To implement a RNDR cycle with the SiSeRO device on a MIT-LL CCID-93 CCD, we clock the output gate (OG) and summing well (SW) regions around the SiSeRO transistor in order to move the charge packet out of the internal gate and store it temporarily under the SW gate. Figure \ref{fig:RNDR_27_waveform} shows the resulting waveform for 57 cycles of RNDR. For this measurement we opted for a single read pixel rate of 650 kHz or $\sim$1.5 \unit{\micro s} per cycle.  The STA Archon CCD controller we use is not able to generate an image from this data out of the box (currently) and we used a combination of raw waveform capture and averaged baseline and signal data with an offline analysis script to produce the image and its corresponding spectrum. The resulting noise performance is shown in Figure \ref{fig:RNDR_27_noise}. A read noise of a single electron is achieved after 13 cyles or $\sim$20 \unit{\micro s}.  Further details of these exciting results can be found in Chattopadhyay et al. 2024 [\citenum{tanmoyspie2024}].

\subsection{Event Processing - with modern tools}
\label{sec:EVNT_PROC}
\begin{figure} [ht]
   \begin{center}
   \begin{tabular}{c}
   \includegraphics[height=6cm]{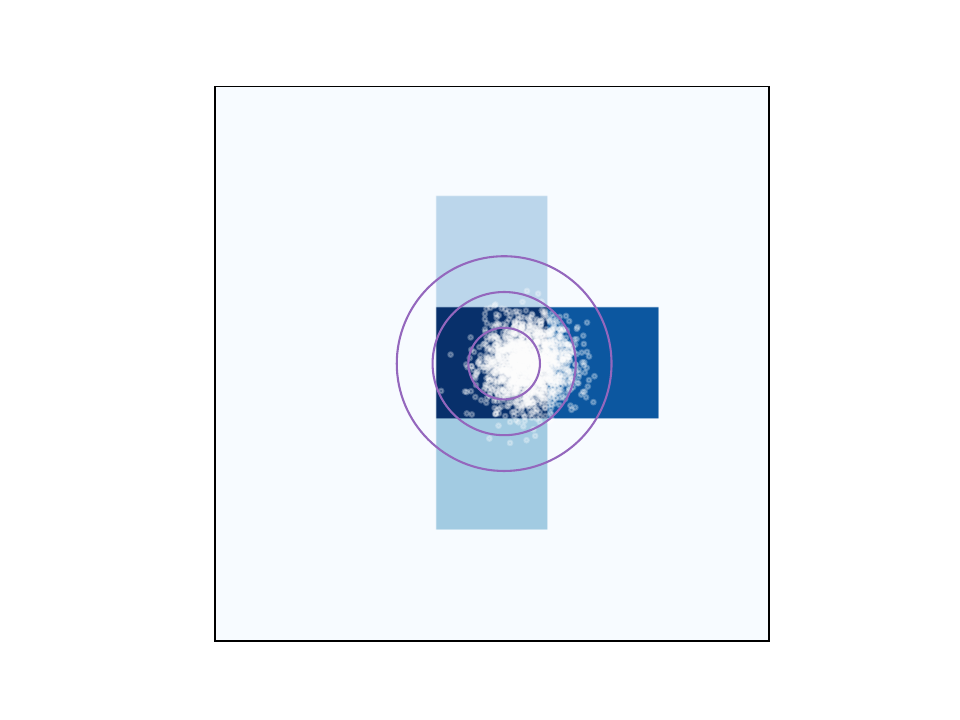}
   \end{tabular}
   \end{center}
   \caption[example] 
   {\label{fig:event5x5} 
Illustration of the Gaussian fitting event reconstruction method for a 6.4 keV photon in an X-ray CCD with 24 $\mu$m pixels of 100 $\mu$m thickness. White dots show the location of individual electrons after they have diffused to the plane within the detector holding the readout gates (using the analytic diffusion model, showing positions before the electric field draws the electrons into the gate itself). Shading denotes the value recorded in each pixel (i.e. the number of electrons falling within the pixel boundary, representing the actual data available from the detector). Purple contours show the best-fitting two-dimensional Gaussian model for the electron cloud, fit via least-squares minimisation to the values recorded in the pixels (denoted by shading). Lines represent the 1, 2 and $3\sigma$ radii, respectively.}
\end{figure}

Advancements in readout hardware are only one way to truly improve the performance of future X-ray missions. The other important path are the algorithms used for the processing of the X-ray events, once the detector is read out. Current algorithms are mostly based on established efforts from the 90s with few upgrades or additions. 
The recent advances in machine learning and large data analysis have generated new interest and promising opportunities to apply these techniques to X-ray telescopes which we are collaborating with our partners at MIT, Smithsonian Astrophysical Observatory (SAO) and Penn State University (PSU).  
One of the areas we are working on is improved event identification and reconstruction from the raw image data the readout system generates. Figure \ref{fig:event5x5} shows an example from simulations of photons absorbed in a typical X-ray CCD. For low energy photon events (especially below 1 keV), charge diffusion results in a significant fraction of the electrons produced in each event diffusing into pixels that do not reach the split threshold.  This results in a shift in the energy scale and reduction in sensitivity to the lowest energy photons when we reconstruct event energies just by summing pixels above threshold. We can recover this lost part of the signal by instead fitting a 2D Gaussian model to the signal recorded in the pixels above threshold to account for the lost electron signals and improving the soft X-ray calibration. Details of this effort can be found in Wilkins et al. 2024 [\citenum{DanAIspie2024}].

\begin{figure} [ht]
   \begin{center}
   \begin{tabular}{c}
   \includegraphics[height=7cm]{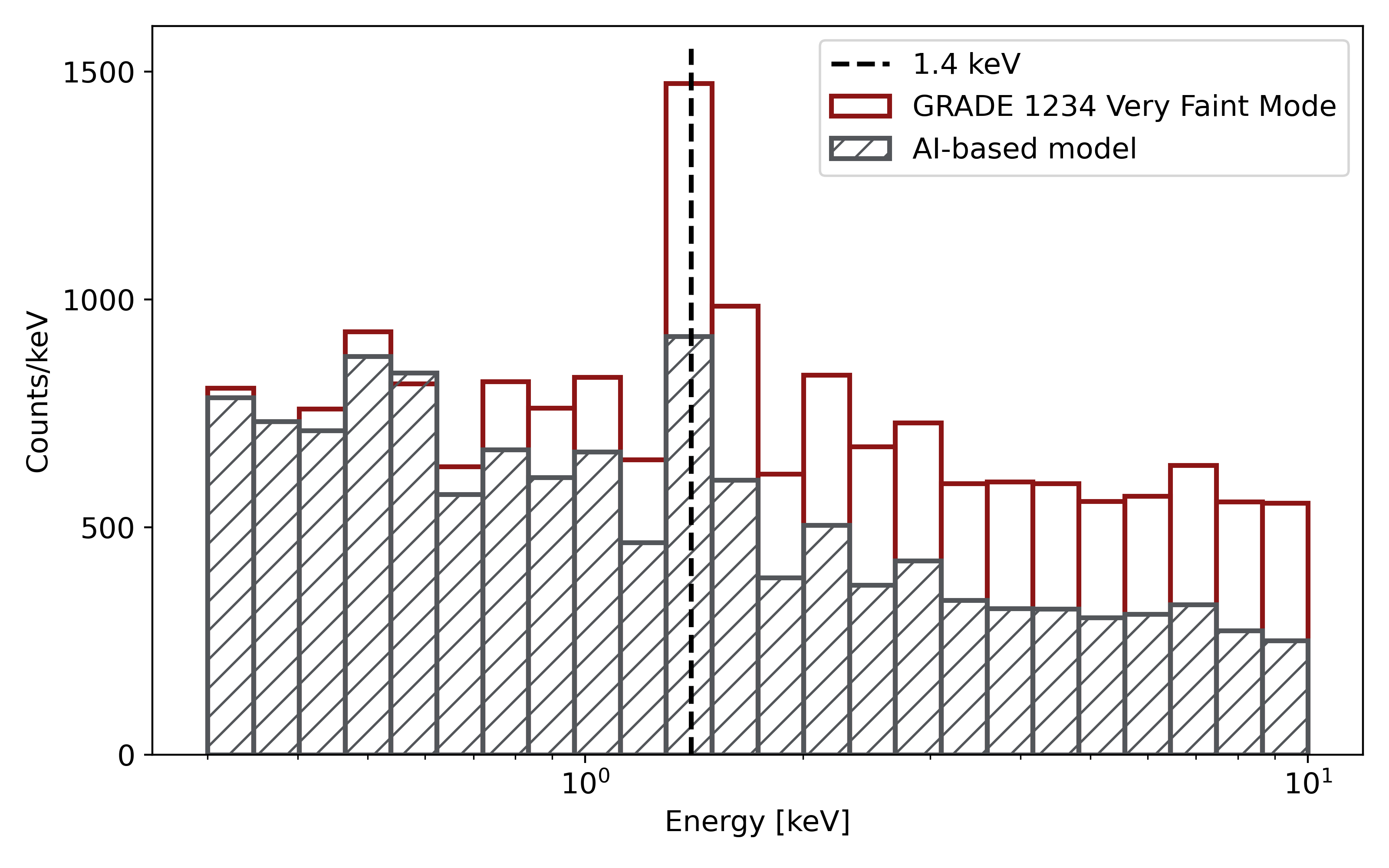}
   \end{tabular}
   \end{center}
   \caption[example] 
   {\label{fig:background} 
AI-based prototype for the reduction of cosmic-ray background allows for $\sim$40\% of relative improvement in cosmic-ray removal over grading algorithms. At the same time it keeps the valid photon rejection level to just 1-2\%, similarly to grading algorithms. Histograms were created after processing a simulated test set with a total of 840000 frames for a Athena WFI like detector. }
\end{figure}

Another field for improved algorithms is the application of machine learning to separate astrophysical X-ray events from radiation background in the detector data. We  have developed prototype machine learning algorithms to identify valid X-ray and cosmic-ray induced background events, trained and tested upon a suite of realistic end-to-end simulations that trace the interaction of cosmic ray particles and their secondaries through the spacecraft and detector. The most recent results of this effort \cite{DanAIspie2024, ArtemMLspie2024} can be seen in Figure \ref{fig:background}: a 40\% relative improvement in cosmic-ray removal over grading algorithms while at the same time keeping the valid photon rejection level to just 1-2\%, similarly to conventional grading algorithms. 

\FloatBarrier
\section{Summary and future plans}
\label{sec:summary}

The Stanford X-ray Astronomy and Observational Cosmology (XOC) group is helping to advance state-of-the-art silicon X-ray imaging technology for future satellite missions.
Together with the Max Planck Institute for Extraterrestrial Physics, we are collaborating on the development of the VERITAS readout ASIC for the Wide Field Imager that will fly on the ESA flagship Athena mission. For nearer-term NASA mission concepts such as AXIS, we are working with MIT/MIT-LL on the development of the MCRC-V1 ASIC as the standard readout approach for multichannel MIT-LL X-ray CCDs, and leveraging advanced digital waveform signal processing to help enable low noise at high speeds. In collaboration with our partners, we are also advancing the novel SiSeRO detector technology, which offers the potential for even greater performance, including demonstrated sub-electron noise through repetitive non-destructive readout. Finally, to leverage fully all aspects of detector performance, we are working with our partners at MIT, the Smithsonian Astrophysical Observatory and Penn State University on the development of advanced event processing algorithms that can provide substantial reductions in the cosmic-ray induced detector background and improvements in the spectroscopic performance at soft X-ray energies.

\acknowledgments 
This work has been supported by NASA APRA grants 80NSSC19K0499 and 80NSSC22K1921, SAT grant 80NSSC23K0211, and the NASA contribution to the Athena WFI via 80GSFC21C0005. We also thank the Kavli Institute of Particle Astrophysics and Cosmology for support via a KIPAC decadal funding grant.

\clearpage

\begin{thebibliography}{10}

\bibitem{AXIS2023}
Reynolds, C.~S., Kara, E.~A., Mushotzky, R.~F., Ptak, A., Koss, M.~J., Williams, B.~J., Allen, S.~W., Bauer, F.~E., Bautz, M., Bogadhee, A., Burdge, K.~B., Cappelluti, N., Cenko, B., Chartas, G., Chan, K.-W., Corrales, L., Daylan, T., Falcone, A.~D., Foord, A., Grant, C.~E., Habouzit, M., Haggard, D., Herrmann, S., Hodges-Kluck, E., Kargaltsev, O., King, G.~W., Kounkel, M., Lopez, L.~A., Marchesi, S., McDonald, M., Meyer, E., Miller, E.~D., Nynka, M., Okajima, T., Pacucci, F., Russell, H.~R., Safi-Harb, S., Strassun, K.~G., Falc{\~a}o, A.~T., Walker, S.~A., Wilms, J., Yukita, M., and Zhang, W.~W., ``{Overview of the advanced x-ray imaging satellite (AXIS)},'' in [{\em UV, X-Ray, and Gamma-Ray Space Instrumentation for Astronomy XXIII}{\nolinebreak\hspace{0.1em}]},  Siegmund, O.~H. and Hoadley, K., eds.,  {\bf 12678},  126781E, International Society for Optics and Photonics, SPIE (2023).

\bibitem{Athena17}
Barcons, X., Barret, D., Decourchelle, A., den Herder, J., Fabian, A., Matsumoto, H., Lumb, D., Nandra, K., Piro, L., Smith, R., and Willingale, R., ``Athena: Esa's x-ray observatory for the late 2020s,'' {\em Astronomische Nachrichten}~{\bf 338},  153--158 (2017).

\bibitem{NOM_WFI_2016}
Meidinger, N., Eder, J., Eraerds, T., Nandra, K., Pietschner, D., Plattner, M., Rau, A., and Strecker, R., ``{The wide field imager instrument for Athena},'' in [{\em Space Telescopes and Instrumentation 2016: Ultraviolet to Gamma Ray}{\nolinebreak\hspace{0.1em}]},  den Herder, J.-W.~A., Takahashi, T., and Bautz, M., eds.,  {\bf 9905},  99052A, International Society for Optics and Photonics, SPIE (2016).

\bibitem{LAZI_DEPFET_2022}
Andricek, L., Bähr, A., Lechner, P., Ninkovic, J., Richter, R., Schopper, F., and Treis, J., ``Depfet—recent developments and future prospects,'' {\em Frontiers in Physics}~{\bf 10} (06 2022).

\bibitem{ericspie2023}
Miller, E.~D., Bautz, M.~W., Grant, C.~E., Foster, R., LaMarr, B., Malonis, A., Prigozhin, G., Schneider, B., Leitz, C., Herrmann, S., Allen, S.~W., Chattopadhyay, T., Orel, P., Morris, G.~R., Stueber, H., Falcone, A.~D., Ptak, A., and Reynolds, C., ``{The high-speed x-ray camera on AXIS},'' in [{\em UV, X-Ray, and Gamma-Ray Space Instrumentation for Astronomy XXIII}{\nolinebreak\hspace{0.1em}]},  Siegmund, O.~H. and Hoadley, K., eds.,  {\bf 12678},  1267816, International Society for Optics and Photonics, SPIE (2023).

\bibitem{2019JATIS...5b1001G}
{Gaskin}, J.~A., {Swartz}, D.~A., {Vikhlinin}, A., {{\"O}zel}, F., {Gelmis}, K.~E., {Arenberg}, J.~W., {Bandler}, S.~R., {Bautz}, M.~W., {Civitani}, M.~M., {Dominguez}, A., {Eckart}, M.~E., {Falcone}, A.~D., {Figueroa-Feliciano}, E., {Freeman}, M.~D., {G{\"u}nther}, H.~M., {Havey}, K.~A., {Heilmann}, R.~K., {Kilaru}, K., {Kraft}, R.~P., {McCarley}, K.~S., {McEntaffer}, R.~L., {Pareschi}, G., {Purcell}, W., {Reid}, P.~B., {Schattenburg}, M.~L., {Schwartz}, D.~A., {Schwartz}, E.~D., {Tananbaum}, H.~D., {Tremblay}, G.~R., {Zhang}, W.~W., and {Zuhone}, J.~A., ``{Lynx X-Ray Observatory: an overview},'' {\em Journal of Astronomical Telescopes, Instruments, and Systems}~{\bf 5},  021001 (Apr. 2019).

\bibitem{Porro_VERITAS2_2014}
Porro, M., Bianchi, D., Vita, G.~D., Herrmann, S., Wassatsch, A., B{\"a}hr, A., Bergbauer, B., Meidinger, N., Ott, S., and Treis, J., ``{VERITAS 2.0 a multi-channel readout ASIC suitable for the DEPFET arrays of the WFI for Athena},'' in [{\em Space Telescopes and Instrumentation 2014: Ultraviolet to Gamma Ray}{\nolinebreak\hspace{0.1em}]},  Takahashi, T., den Herder, J.-W.~A., and Bautz, M., eds.,  {\bf 9144},  91445N, International Society for Optics and Photonics, SPIE (2014).

\bibitem{SchweingruberSPIE2024}
Schweingruber, A., Allen, S., Albrecht, S., Herrmann, S., Mayr, A., Morris, G., Mueller-Seidlitz, J., Nandra, K., Orel, P., Reifers, J., Schnetler, H., and Dakshinamurthy, A.~K., ``{The VERITAS 2.3 readout ASIC for the ATHENA Wide Field Imager},'' in [{\em Space Telescopes and Instrumentation 2024: Ultraviolet to Gamma Ray}{\nolinebreak\hspace{0.1em}]},   {\bf 13093},  13093164, International Society for Optics and Photonics, SPIE (2024).

\bibitem{porelMCRCspie2022}
Orel, P., Herrmann, S., Chattopadhyay, T., Morris, G.~R., Allen, S.~W., Prigozhin, G.~Y., Foster, R., Malonis, A., Bautz, M.~W., Cooper, M.~J., and Donlon, K., ``{X-ray speed reading with the MCRC: a low noise CCD readout ASIC enabling readout speeds of 5 Mpixel/s/channel},'' in [{\em X-Ray, Optical, and Infrared Detectors for Astronomy X}{\nolinebreak\hspace{0.1em}]},  Holland, A.~D. and Beletic, J., eds.,  {\bf 12191},  1219124, International Society for Optics and Photonics, SPIE (2022).

\bibitem{porelMCRCspie2024}
Orel, P., Herrmann, S., Chattopadhyay, T., Morris, G.~R., Stueber, H., Pan, A., Allen, S.~W., Wilkins, D., Prigozhin, G.~Y., LaMarr, B., Foster, R., Malonis, A., Bautz, M.~W., Cooper, M.~J., and Donlon, K., ``{X-ray speed reading with the MCRC: prototype success and next generation upgrades},'' in [{\em X-Ray, Optical, and Infrared Detectors for Astronomy XI}{\nolinebreak\hspace{0.1em}]},   {\bf 13103},  1310332, International Society for Optics and Photonics, SPIE (2024).

\bibitem{stueberSPIE2024}
Stueber, H.~R., Chattopadhyaya, T., Herrmann, S.~C., Orel, P., Gebre, T., Joshi, A., Allen, S.~W., Morris, G., and Poliszczuk, A., ``{The XOC X-ray Beamline: Probing Colder, Quieter, and Softer},'' in [{\em X-Ray, Optical, and Infrared Detectors for Astronomy XI}{\nolinebreak\hspace{0.1em}]},   {\bf 13103},  13103--77, International Society for Optics and Photonics, SPIE (2024).

\bibitem{chattopadhyay20}
Chattopadhyay, T., Herrmann, S., Allen, S.~W., Hirschman, J., Morris, G., Bautz, M., Malonis, A., Foster, R., Prigozhin, G., Craig, D., and Burke, B., ``{Tiny-box: a tool for the versatile development and characterization of low noise fast x-ray imaging detectors},'' in [{\em X-Ray, Optical, and Infrared Detectors for Astronomy IX}{\nolinebreak\hspace{0.1em}]},   {\bf 11454},  1145423, International Society for Optics and Photonics, SPIE (2020).

\bibitem{chattopadhyay23_sisero_digfilt}
{Chattopadhyay}, T., {Herrmann}, S., {Kaplan}, M., {Orel}, P., {Donlon}, K., {Prigozhin}, G., {Morris}, G., {Cooper}, M., {Malonis}, A., {Allen}, S.~W., {Bautz}, M.~W., and {Leitz}, C., ``{Improved noise performance from the next-generation buried-channel p-MOSFET SiSeROs},'' {\em Journal of Astronomical Telescopes, Instruments, and Systems}~{\bf 9},  026001 (Apr. 2023).

\bibitem{tiffenberg17}
Tiffenberg, J., Sofo-Haro, M., Drlica-Wagner, A., Essig, R., Guardincerri, Y., Holland, S., Volansky, T., and Yu, T.-T., ``Single-electron and single-photon sensitivity with a silicon skipper ccd,'' {\em Phys. Rev. Lett.}~{\bf 119},  131802 (Sep 2017).

\bibitem{RNDR_DEPFET_2007}
Wolfel, S., Herrmann, S., Lechner, P., Lutz, G., Porro, M., Richter, R.~H., Struder, L., and Treis, J., ``A novel way of single optical photon detection: Beating the 1/f noise limit with ultra high resolution depfet-rndr devices,'' {\em IEEE Transactions on Nuclear Science}~{\bf 54}(4),  1311--1318 (2007).

\bibitem{sofo23_nsisero}
{Sofo-Haro}, M., {Donlon}, K., {Estrada}, J., {Holland}, S., {Fahim}, F., and {Leitz}, C., ``{Achieving Single-Electron Sensitivity at Enhanced Speed in Fully-Depleted CCDs with Double-Gate MOSFETs},'' {\em arXiv e-prints} ,  arXiv:2310.13644 (Oct. 2023).

\bibitem{kemmer87_depfet}
Kemmer, J. and Lutz, G., ``New detector concepts,'' {\em Nuclear Instruments and Methods in Physics Research Section A: Accelerators, Spectrometers, Detectors and Associated Equipment}~{\bf 253}(3),  365--377 (1987).

\bibitem{strueder00_depfet_imager}
Strueder, L., Meidinger, N., Pfeffermann, E., Hartmann, R., Braeuninger, H.~W., Krause, N., Hartner, G.~D., Dennerl, K., Haberl, F., Kemmer, S., Popp, M., Truemper, J.~E., Kollmer, J., Johannes, T., Lutz, G., Hauff, D., Richter, R.~H., Klein, P., Hoernel, N., Solc, P., Eckhardt, R., Fischer, P., Neeser, W., Ulrici, J., Wermes, N., Holl, P., Lechner, P., Kemmer, J., Soltau, H., Stoetter, R., Weber, U., and Weichert, U., ``{Fully depleted backside-illuminated spectroscopic active pixel sensors from the infrared to x rays (1 eV to 25 keV)},'' in [{\em X-Ray Optics, Instruments, and Missions III}{\nolinebreak\hspace{0.1em}]},  Truemper, J.~E. and Aschenbach, B., eds.,  {\bf 4012},  200 -- 217, International Society for Optics and Photonics, SPIE (2000).

\bibitem{treberspurg20_wfi}
{Treberspurg}, W., {Andritschke}, R., {Behrens}, A., {Bonholzer}, M., {Emberger}, V., {Hauser}, G., {Lechner}, P., {Meidinger}, N., and {M{\"u}ller-Seidlitz}, J., ``{Characterization of a 256 {\texttimes} 256 pixel DEPFET detector for the WFI of Athena},'' {\em Nuclear Instruments and Methods in Physics Research A}~{\bf 958},  162555 (Apr. 2020).

\bibitem{matsunaga91}
Matsunaga, Y., Yamashita, H., and Ohsawa, S., ``A highly sensitive on-chip charge detector for ccd area image sensor,'' {\em IEEE Journal of Solid-State Circuits}~{\bf 26}(4),  652--656 (1991).

\bibitem{chattopadhyay22_sisero}
Chattopadhyay, T., Herrmann, S., Burke, B.~E., Donlon, K., Prigozhin, G., Morris, G., Orel, P., Cooper, M., Malonis, A., Wilkins, D.~R., Suntharalingam, V., Allen, S.~W., Bautz, M.~W., and Leitz, C., ``{First results on SiSeRO devices: a new x-ray detector for scientific instrumentation},'' {\em Journal of Astronomical Telescopes, Instruments, and Systems}~{\bf 8}(2),  1 -- 12 (2022).

\bibitem{rodrigues21}
{Rodrigues}, D., {Andersson}, K., {Cababie}, M., {Donadon}, A., {Botti}, A., {Cancelo}, G., {Estrada}, J., {Fernandez-Moroni}, G., {Piegaia}, R., {Senger}, M., {Haro}, M.~S., {Stefanazzi}, L., {Tiffenberg}, J., and {Uemura}, S., ``{Absolute measurement of the Fano factor using a Skipper-CCD},'' {\em Nuclear Instruments and Methods in Physics Research A}~{\bf 1010},  165511 (Sept. 2021).

\bibitem{tanmoyspie2024}
Chattopadhyay, T., Herrmann, S.~C., Orel, P., Donlon, K., Allen, S.~W., Bautz, M.~W., Cantrall, B.~J., Cooper, M.~J., LaMarr, B.~J., Leitz, C.~W., Miller, E.~D., Morris, G.~R., Prigozhin, G.~Y., Prigozhin, I., Stueber, H.~R., and Wilkins, D.~R., ``{Demonstrating sub-electron noise performance in SiSeRO (Single electron Sensitive Readout) devices},'' in [{\em X-Ray, Optical, and Infrared Detectors for Astronomy XI}{\nolinebreak\hspace{0.1em}]},   {\bf 13103},  1310357, International Society for Optics and Photonics, SPIE (2024).

\bibitem{DanAIspie2024}
Wilkins, D., Poliszczuk, A., Schneider, B., Allen, S., Miller, E., Bautz, M., Chattopadhyay, T., Falcone, A., Foster, R., Grant, C., Herrmann, S., Morris, G., Nulsen, P., Orel, P., and Schellenberger, G., ``{Augmenting astronomical X-ray detectors with AI for enhanced sensitivity and reduced background},'' in [{\em Space Telescopes and Instrumentation 2024: Ultraviolet to Gamma Ray}{\nolinebreak\hspace{0.1em}]},   {\bf 13093},  13093--65, International Society for Optics and Photonics, SPIE (2024).

\bibitem{ArtemMLspie2024}
Poliszczuk, A., Wilkins, D., Allen, S., Miller, E., Chattopadhyay, T., Schneider, B., Darve, J., Bautz, M., Falcone, A., Foster, R., Grant, C., Herrmann, S., Kraft, R., Morris, G., Nulsen, P., Orel, P., Schellenberger, G., and Stueber, H., ``{Towards efficient machine-learning-based reduction of the cosmic-ray induced background in X-ray imaging detectors: increasing context awareness},'' in [{\em Space Telescopes and Instrumentation 2024: Ultraviolet to Gamma Ray}{\nolinebreak\hspace{0.1em}]},   {\bf 13093},  13093--65, International Society for Optics and Photonics, SPIE (2024).

\end{thebibliography}

\end{document}